\documentclass{aip-cp}

\usepackage[numbers]{natbib}
\usepackage{rotating}
\usepackage{graphicx}
\usepackage{lineno}

\begin{document}

%% Mine:
\newcommand{\nonubb}  {$0 \nu \beta \beta$}
\newcommand{\twonubb}{$2 \nu \beta \beta$}
\newcommand{\bb}{$\beta\beta$}
\newcommand{\onecpRty} {\hbox{1~count/ROI/t-y}}
\newcommand{\MJ}{{\sc{Majo\-ra\-na}}}
\newcommand{\Dem}{{\sc{Demonstrator}}}
\newcommand{\MJD}{{\sc{Majorana Demonstrator}}}
\def\ppc{P-PC}  
\newcommand{\gam}{$\gamma$}
\def\nuc#1#2{${}^{#1}$#2}
\def\cpRty{c/(ROI t yr)}

% Title portion
\title{The Status and Initial Results of the {\sc Majorana Demonstrator} Experiment}

\author[usc]{V.E.~Guiseppe\corref{cor1}}
\author[lbnl]{N.~Abgrall}
\author[uw]{S.I.~Alvis}
\author[pnnl]{I.J.~Arnquist}
\author[usc,ornl]{F.T.~Avignone~III}
\author[ITEP]{A.S.~Barabash}
\author[usd]{C.J.~Barton}
\author[ornl]{F.E.~Bertrand}
\author[mpi]{T.~Bode}
\author[lbnl]{A.W.~Bradley}
\author[JINR]{V.~Brudanin}
\author[duke,tunl]{M.~Busch}
\author[uw]{M.~Buuck}
\author[unc,tunl]{T.S.~Caldwell}
\author[lbnl]{Y-D.~Chan}
\author[sdsmt]{C.D.~Christofferson}
\author[lanl]{P.-H.~Chu}
\author[uw,clara]{C. Cuesta}
\author[uw]{J.A.~Detwiler}
\author[sdsmt]{C. Dunagan}
\author[ut,ornl]{Yu.~Efremenko}
\author[ou]{H.~Ejiri}
\author[lanl]{S.R.~Elliott}
\author[unc,tunl]{T.~Gilliss}
\author[princeton]{G.K.~Giovanetti}
\author[ncsu,tunl,ornl]{M.P.~Green}
\author[uw]{J. Gruszko}
\author[uw]{I.S.~Guinn}
%\author[usc]{V.E.~Guiseppe}
\author[unc,tunl]{C.R.~Haufe}
\author[lbnl]{L.~Hehn}
\author[unc,tunl]{R.~Henning}
\author[pnnl]{E.W.~Hoppe}
\author[unc,tunl]{M.A.~Howe}
\author[blhill]{K.J.~Keeter}
\author[ttu]{M.F.~Kidd}
\author[ITEP]{S.I.~Konovalov}
\author[pnnl]{R.T.~Kouzes}
\author[ut]{A.M.~Lopez}
\author[queens]{R.D.~Martin}
\author[lanl]{R. Massarczyk}
\author[unc,tunl]{S.J.~Meijer}
\author[mpi,tum]{S.~Mertens}
\author[lbnl]{J.~Myslik}
\author[unc,tunl]{C. O'Shaughnessy}
\author[unc,tunl]{G.~Othman}
\author[lbnl]{A.W.P.~Poon}
\author[ornl]{D.C.~Radford}
\author[unc,tunl]{J.~Rager}
\author[unc,tunl]{A.L.~Reine}
\author[lanl]{K.~Rielage}
\author[uw]{R.G.H.~Robertson}
\author[uw]{N.W.~Rouf}
\author[unc,tunl]{B.~Shanks}
\author[JINR]{M.~Shirchenko}
\author[sdsmt]{A.M.~Suriano}
\author[usc]{D.~Tedeschi}
\author[unc,tunl]{J.E.~Trimble}
\author[ornl]{R.L.~Varner}
\author[JINR]{S. Vasilyev}
\author[lbnl]{K.~Vetter}
\author[unc,tunl]{K.~Vorren}
\author[lanl]{B.R.~White}
\author[unc,tunl,ornl]{J.F.~Wilkerson}
\author[usc]{C. Wiseman}
\author[usd]{W.~Xu}
\author[JINR]{E.~Yakushev}
\author[ornl]{C.-H.~Yu}
\author[ITEP]{V.~Yumatov}
\author[JINR]{I.~Zhitnikov}
\author[lanl]{B.X.~Zhu}

\affil[usc]{Department of Physics and Astronomy, University of South Carolina, Columbia, SC, USA}
\affil[lbnl]{Nuclear Science Division, Lawrence Berkeley National Laboratory, Berkeley, CA, USA}
\affil[uw]{Center for Experimental Nuclear Physics and Astrophysics, and Department of Physics, University of Washington, Seattle, WA, USA}
\affil[pnnl]{Pacific Northwest National Laboratory, Richland, WA, USA}
\affil[ornl]{Oak Ridge National Laboratory, Oak Ridge, TN, USA}
\affil[ITEP]{National Research Center ``Kurchatov Institute'' Institute for Theoretical and Experimental Physics, Moscow, Russia}
\affil[usd]{Department of Physics, University of South Dakota, Vermillion, SD, USA} %institution change to just dept of phys. 10/11
\affil[mpi]{Max-Planck-Institut f\"{u}r Physik, M\"{u}nchen, Germany}
\affil[JINR]{Joint Institute for Nuclear Research, Dubna, Russia}
\affil[duke]{Department of Physics, Duke University, Durham, NC, USA}
\affil[tunl]{Triangle Universities Nuclear Laboratory, Durham, NC, USA}
\affil[unc]{Department of Physics and Astronomy, University of North Carolina, Chapel Hill, NC, USA}
\affil[sdsmt]{South Dakota School of Mines and Technology, Rapid City, SD, USA}
\affil[lanl]{Los Alamos National Laboratory, Los Alamos, NM, USA}
\affil[ut]{Department of Physics and Astronomy, University of Tennessee, Knoxville, TN, USA}
\affil[ou]{Research Center for Nuclear Physics, Osaka University, Ibaraki, Osaka, Japan}
\affil[princeton]{Department of Physics, Princeton University, Princeton, NJ, USA}
\affil[ncsu]{Department of Physics, North Carolina State University, Raleigh, NC, USA}	%, removed as institution: retroactively on 11/24/14 to a date of 11/01/13,added back with Matt's move 8/15
\affil[blhill]{Department of Physics, Black Hills State University, Spearfish, SD, USA}
\affil[ttu]{Tennessee Tech University, Cookeville, TN, USA}
\affil[queens]{Department of Physics, Engineering Physics and Astronomy, Queen's University, Kingston, ON, Canada} %removed as institution 5/12, added back with Ryan's move 9/15
\affil[tum]{Physik Department, Technische Universit\"{a}t, M\"{u}nchen, Germany}
%\affil[uchic}{Department of Physics, University of Chicago, Chicago, IL, USA}
%\affil[ucne}{Department of Nuclear Engineering, University of California, Berkeley, CA, USA} removed as institution 1/11
%\affil[ucph}{Department of Physics, University of California, Berkeley, CA, USA} removed as institution 1/11
%\affil[sjtu}{Shanghai Jiao Tong University, Shanghai, China}, removed as institution: retroactively on 11/24/14 to a date of 11/01/13
\affil[clara]{Present Address: Centro de Investigaciones Energ\'{e}ticas, Medioambientales y Tecnol\'{o}gicas, CIEMAT, 28040, Madrid, Spain}

\corresp[cor1]{Corresponding author: guiseppe@sc.edu}

\maketitle

%\linenumbers

\begin{abstract}
Neutrinoless double-beta decay searches play a major role in determining the nature of neutrinos, the existence of a lepton violating process, and the effective Majorana neutrino mass. The {\sc Majorana} Collaboration assembled an array of high purity Ge detectors to search for neutrinoless double-beta decay in $^{76}$Ge. The {\sc Majorana} {\sc Demonstrator} is comprised of 44.1 kg (29.7 kg enriched in $^{76}$Ge) of Ge detectors  divided between two modules contained in a low-background shield at the Sanford Underground Research Facility in Lead, South Dakota, USA. The initial goals of the {\sc Demonstrator} are to establish the required background and scalability of a Ge-based next-generation ton-scale experiment. Following a commissioning run that started in 2015, the first detector module started low-background data production in early 2016.  The second detector module was added in August 2016 to begin operation of the entire array. We discuss results of the initial physics runs, as well as the status and physics reach of the full  {\sc Majorana} {\sc Demonstrator} experiment. 

\end{abstract}

\section{EXPERIMENTAL OVERVIEW}

The discovery of Majorana neutrinos would have profound
theoretical implications in extending the standard model of particle physics
while yielding insights into the origin of mass itself.
Experimental evidence of neutrinoless double-beta (\nonubb) decay
would definitively reveal the Majorana nature of neutrinos and establish a lepton violating process. 
The \MJ\ Collaboration is pursuing R\&D aimed at a ton-scale,
$^{76}$Ge \nonubb-decay experiment through the operation of  
a \Dem\ array \cite{abg14,ell16}
 on the 4850-foot level of the Sanford Underground Research Facility (SURF) \cite{hei15}  in Lead, SD, USA. The goals are to 1) demonstrate backgrounds are low enough to justify moving towards a ton-scale \nonubb\ experiment, 2) establish the feasibility of constructing and fielding modular arrays of Ge detectors, and 3) search for additional physics beyond the standard model. %The assay and Monte Carlo simulation program project the backgrounds in the \Dem\ to be $\leq3.5$ \cpRty. 
 The background goal in the \nonubb\ peak region of interest (ROI) is 3 \cpRty\ after analysis cuts. 
The \Dem\ contains a total of 44.1 kg of Ge detectors with 29.7 kg of the detector mass enriched to 88\% in the isotope $^{76}$Ge. 
The p-type, point contact detectors \cite{bar07,luk89} in use offer excellent energy resolution, background rejection methods, and low energy thresholds. The detectors are split between two separate modules (1 and 2) that are surrounded by a low-background passive shield with active muon veto.

\subsection{Material Purity and Background Rejection}
A key to the \MJ\ design is the strict material purity requirements for parts used in the construction of the array. Prior to construction, a comprehensive assay program sampled candidate materials in order to identify acceptable detector components. A full Monte Carlo simulation of the \MJD\ geometry converts the assay radio-purity levels or limits into a total background projection of $\leq3.5$ \cpRty. A listing of all simulated background contributions can be found in Ref. \cite{abg16a}. 
%Naturally occurring U and Th radioactivity in the materials close to the detectors present the greatest material selection challenge. 
In addition to its excellent thermal and mechanical properties, the purity of underground electroformed Cu makes it the ideal choice for the components in proximity to the detector:  the detector unit and string assemblies, the cryostat, and the innermost passive shield. Our underground electrofoming baths 
%(10 at SURF and 6 at PNNL) 
produced 2474 kg of ultra pure Cu ($\leq 0.1 \mu$Bq/kg Th and $\leq 0.1 \mu$Bq/kg U \cite{abg16a}).
%%%%reword
All Cu and plastics parts were fabricated in a dedicated underground machine shop at SURF, cleaned in an adjacent wet-lab cleanroom, and tracked through a custom database \cite{orr15} during assembly of the array.  
%A dedicated Cu machine shop operated underground at SURF fabricated all Cu and plastic parts and an adjacent wet-lab cleanroom allowed for a thorough cleaning of all parts used during assembly of the array. 
A validation of cleaning procedures ensured clean parts remained clean through the handling and processing of parts. 
The remaining nearby components went through a careful selection and design process to ensure low mass and acceptable purity. Custom front-end boards were designed and fabricated to house the first stage of detector readout electronics. Fine coaxial cables were selected for the signal and high voltage wiring using custom miniature connectors. With the innermost passive shield layer made from underground electroformed Cu, the additional layers of the passive shield contain commercially sourced Cu and Pb that also underwent an aggressive surface cleaning procedure to remove contaminants from fabrication and handling.

%rejection
The background rejection strategies of the \MJD\ start with the Ge detectors. The process of enrichment, zone refining, and crystal pulling provides the  necessary purification of the Ge metal from unwanted radioactive isotopes. The cosmogenic activation of the purified enriched Ge was controlled through underground storage and shielding during all stages of Ge detector fabrication \cite{abg17b}. The excellent energy resolution (2.4 keV FWHM at 2039 keV) of of a p-type, point-contact Ge detector allows discrimination of the \twonubb\ spectrum of events from the narrow \nonubb\ region. Further, the slow drift velocity and localized weighting potential allows separation in time of multi-site events occurring in a detector due to gamma-ray interactions.  The short range of the emitted electrons allows double-beta decay to be characterized as a single-site event. This multi-site event cut is achieved through Pulse Shape Discrimination (PSD) of the digitized waveform  by comparing the maximum Amplitude (A) of the current pulse to the reconstructed Energy (E) through the calculation of an `AvsE' parameter \cite{bud09}. 
%Smaller AvsE parameters are more indicative of a multi-site background event. 
Alpha-emitting contamination on the surface of the Ge detector or nearby materials contribute a background that can be identified by a signature detector response. An alpha striking the passivated surface of the Ge detector results in prompt collection of a fraction of the alpha energy while the remainder is collected more slowly. Due to their separate time components, only the fast component is reconstructed as the event energy. A Delayed Charge Recovery (DCR) parameter \cite{gru16} relates the greater slope of the waveform tail to the slow charge collection component and the presence of a surface alpha event. 
Additional background suppression is achieved through a passive shield containing Cu and Pb 
%containing 5 cm of underground electroformed Cu, 5 cm of commercially sourced Cu, and 45 cm of low background Pb. 
within a sealed aluminum enclosure that defines a nitrogen gas purge volume to displace radon gas. Two layers of plastic scintillating panels tag penetrating muons to enable an active veto of prompt muon-induced backgrounds. Finally, a layer of high density polyethylene aids in moderating neutrons. 

\subsection{Operation and Data Sets}
The \MJD\ has been operating with enriched Ge detectors since May 2015. Changes in configuration or operational status are marked by distinct Data Sets (DS). 
%Only some of the currently analyzed data sets are reported here. 
Module 1 (M1) occupied the shield prior to the installation of the inner electroformed Cu shield for DS-0. DS-1 (operated in a data prescaling blindness mode between Dec. 31, 2015 - May 24, 2016) achieved lower backgrounds due to the presence of the innermost shield. The background levels of DS-0 and DS-1 are reported in Ref. \cite{ell16}. Analysis is ongoing for M1 within DS-2 (May 24 - July 14, 2016) where multi-sampling is enabled to take a longer waveform to improve the DCR cut. With the addition of Module 2 (M2)  to the shield, the results within the DS-3 and DS-4 full-array operation (as separate data streams) from Aug. 25 - Sep. 27, 2016 are presented here. Since that time, the array has been collecting physics data within DS-5 (as a single data stream) and DS-6 (a return to blindness mode). Analysis is ongoing over all data sets with a combined background and \nonubb\ limit to be released soon. 

\section{INITIAL RESULTS}

\begin{figure}[t]
\centering
\begin{tabular}[b]{c}
  \includegraphics[width=0.32\textwidth]{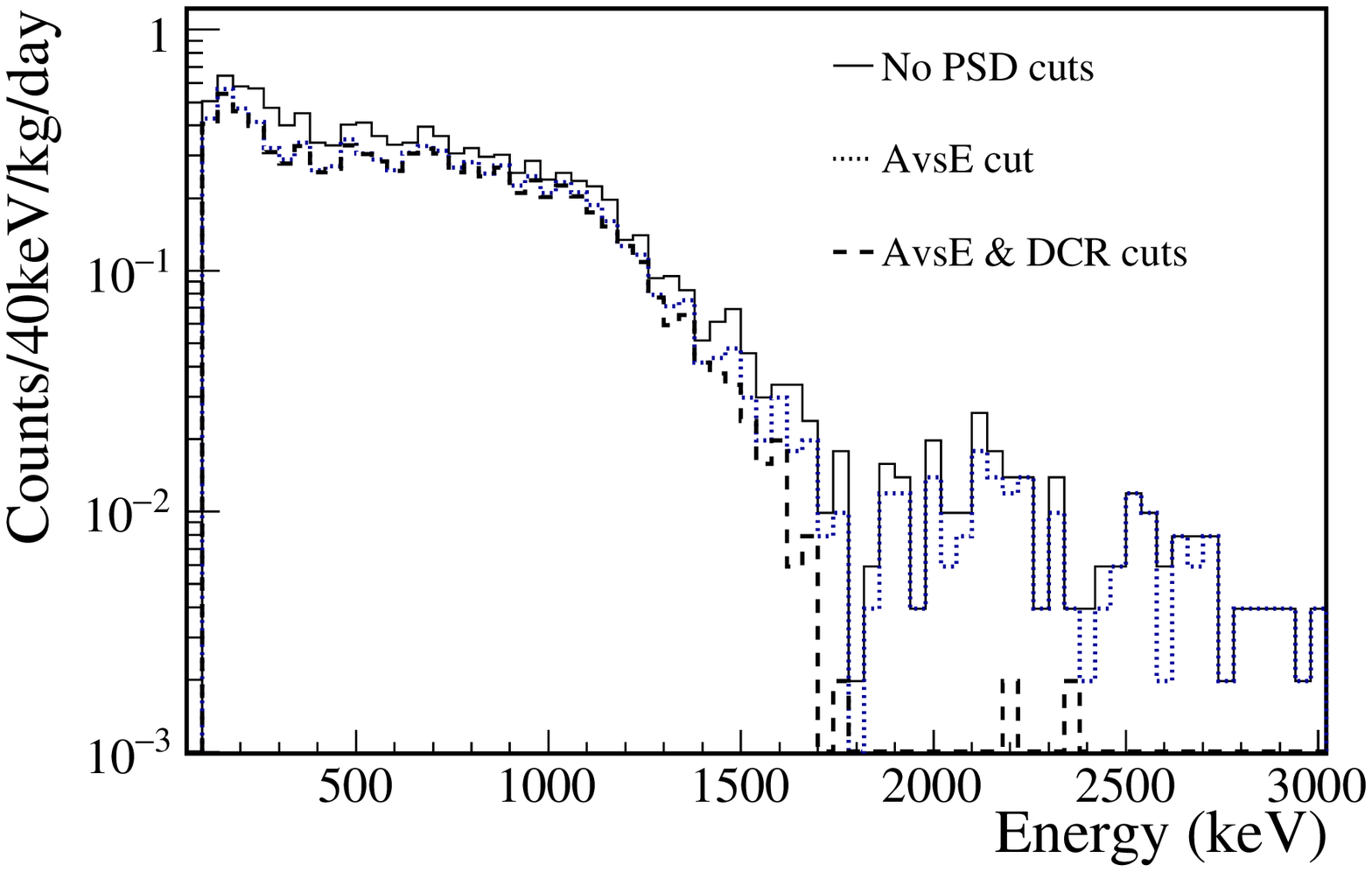} \\
  \small (a)
\end{tabular}
% \quad
\begin{tabular}[b]{c}
  \includegraphics[width=0.32\textwidth]{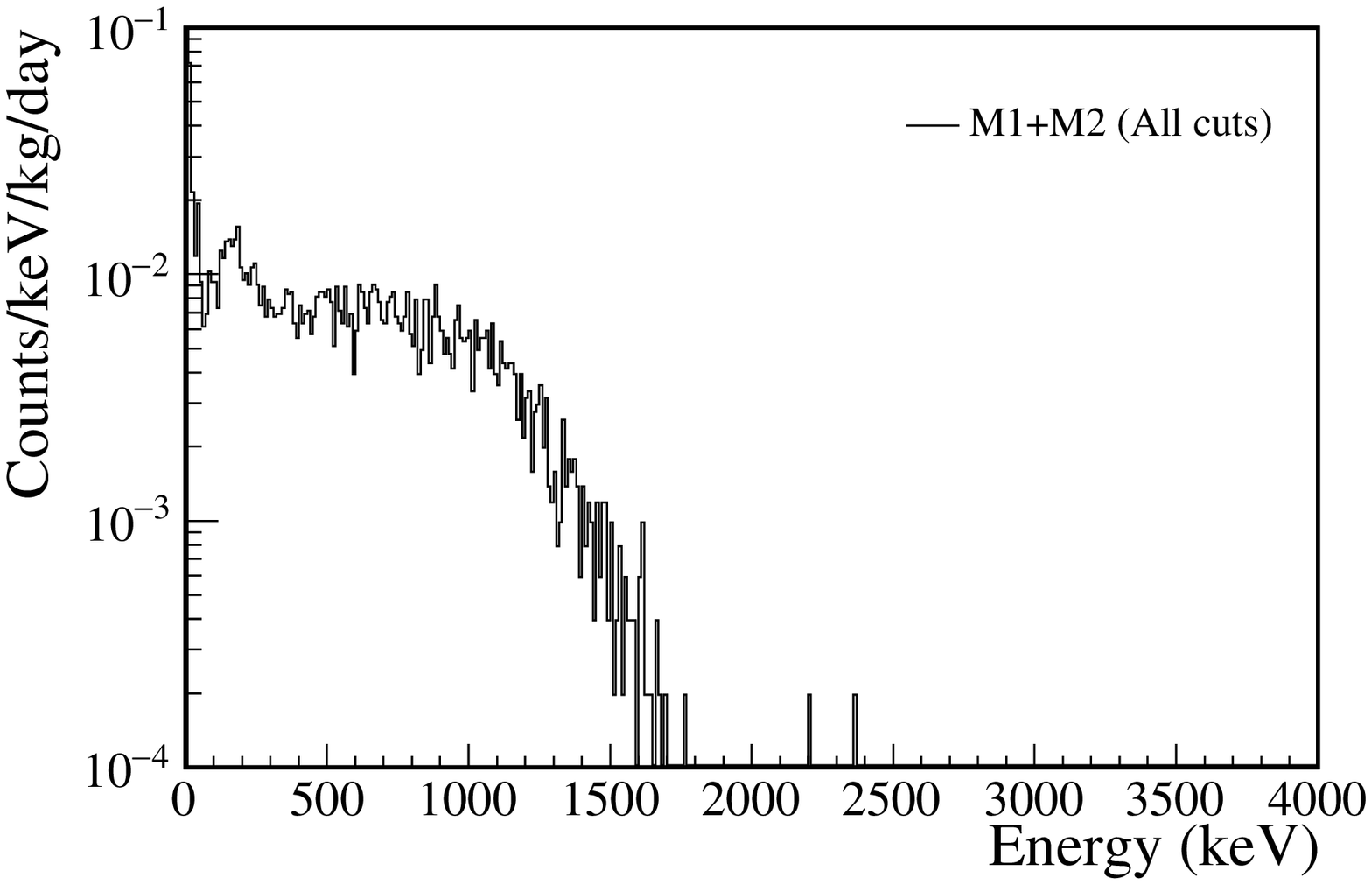}\\
  \small (b) 
  \end{tabular}
  \begin{tabular}[b]{c}
  \includegraphics[width=0.32\textwidth]{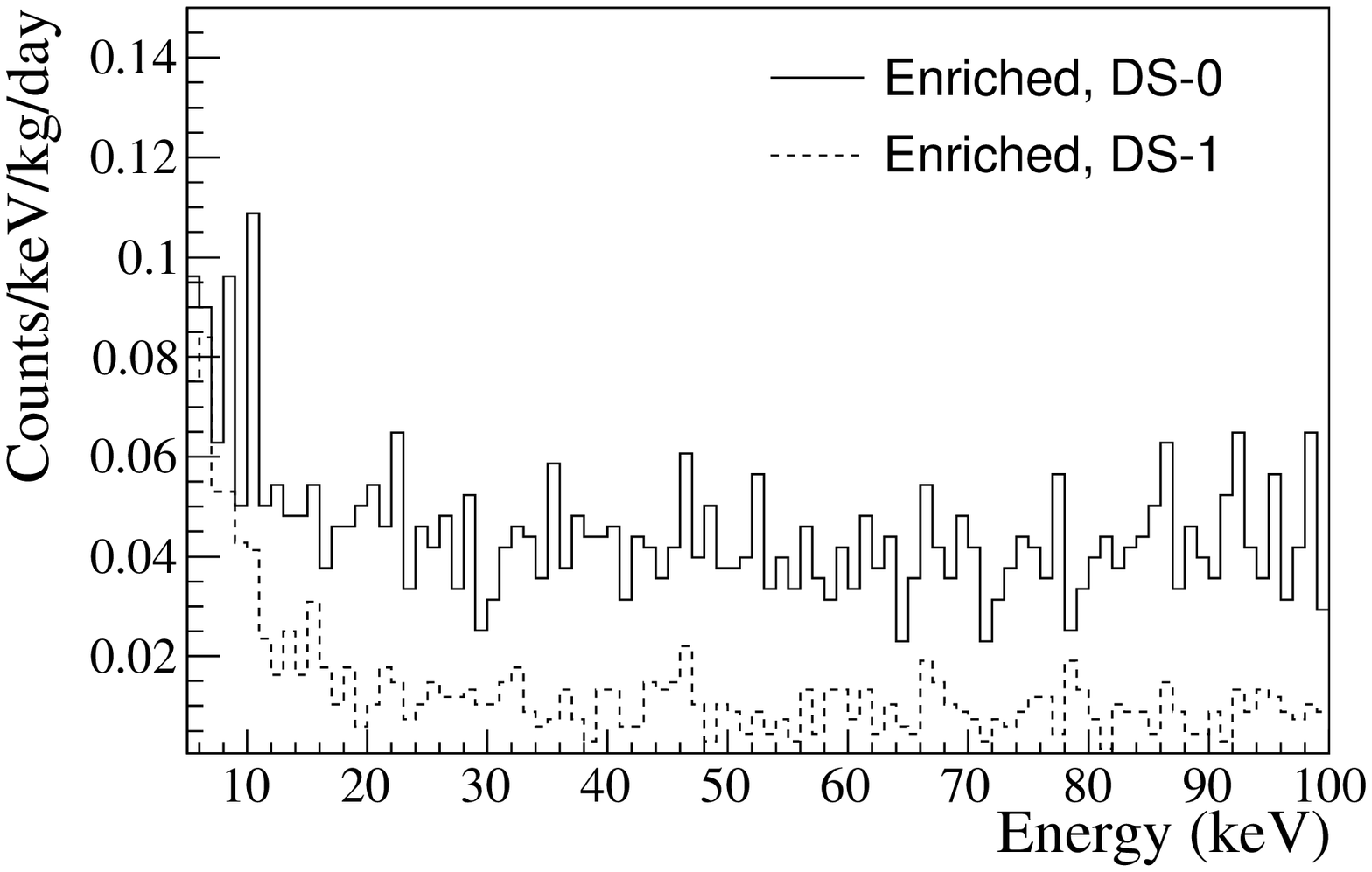} \\
  \small (c) 
  \end{tabular}
  \caption{(a) Enriched detector background spectrum from Data Sets 3 and 4 (an exposure of 1.39 kg$\cdot$y) before and after Pulse Shape Discrimination (PSD) cuts. The two cuts applied are the Amplitude vs. Energy (AvsE) multisite event rejection and the Delayed Charge Recovery (DCR) surface alpha cut. (b) The same enriched detector background spectrum after all analysis cuts at a finer energy binning. 
  %Based on a 400-keV window centered at the \nonubb energy peak at 2029 keV, the background index of 1.8$\times 10^{-3}$ \cpRty.
 (c) The low energy spectrum from DS-0 and DS-1 after initial data cleaning cuts. The improvement in DS-1 is due to the final configuration of the passive shield.}
  \label{fig:DS3DSCuts}
\end{figure}

The first analysis of data from DS-0 and DS-1 M1 data are reported in Ref. \cite{ell16}. Since that release, the Collaboration has been improving the PSD algorithms and added the second module. The latest analysis of data is from 1.39 kg$\cdot$y of enriched exposure in DS-3 \& DS-4, which contain both modules operating in the low background shield configuration. The energy spectrum, after data cleaning cuts, is displayed in Fig. \ref{fig:DS3DSCuts}(a) along with the sequential effect of performing the AvsE cut to remove multiple site events, such as background gamma rays, and the DCR cut to remove surface alpha backgrounds. The resulting energy spectrum (in Fig. \ref{fig:DS3DSCuts}(b) at a finer energy binning) leaves one main feature: the \twonubb\ spectrum. Though the \nonubb\ region of interest is $<3$-keV wide centered at 2039 keV, a background index determination is made from a larger 400 keV window on the same center. The projected background rate is 5.1$^{+8.9}_{-3.2}$ \cpRty\ where the ROI is pinned at 2.9 keV  and 2.6 keV, respectively, for M1 and M2 based on the average detector energy resolutions. The corresponding background index is $1.8\times10^{-3}$ c/(keV kg y).

The excellent energy resolution and low energy thresholds of p-type, point contact detectors open up sensitivity to additional low energy physics searches. The controlled surface exposure of the enriched Ge material limited the production of internal cosmogenic activation, which would be the dominant source of backgrounds at low energy in a shielded low-background experiment. 
A low-background, low-energy spectrum allows searches for pseudoscalar dark matter, vector dark matter, solar axions, and other exotic physics beyond the standard model. Our analysis of DS-0 achieved a background rate of around 0.04 c/(kev kg d) near 20 keV with limits on low energy physics reported in Ref. \cite{abg17}. Data cleaning cuts and improved data analysis continue in subsequent data sets where the inner shield offers lower backgrounds. The improved background rate can be seen in Fig. \ref{fig:DS3DSCuts}(c) with a background rate $<0.02$ c/(kev kg d) near 20 keV. The efficiency of the detectors below 5 keV is under study in order to open up the spectrum and cut non-physics events down to the detector threshold.

%\begin{figure}[h]
%\centering
%\begin{tabular}[b]{c}
 % \includegraphics[width=0.45\textwidth]{DS0andDS1-LowE_BW.eps} \\
  %\end{tabular}
  %\caption{The low energy spectrum from DS-0 and DS-1 after initial data cleaning cuts. The improvement in DS-1 is due to the final configuration of the passive shield.}
  %\label{fig:LowE}
%\end{figure}

\section{OUTLOOK}

The \MJ\ collaboration is operating its \Dem\ array as R\&D for a ton-scale \nonubb-decay experiment. Initial analysis shows that we have achieved the best energy resolution (2.4 keV FWHM) of any \nonubb\ experiment. The PSD  provides a reduction of background events leaving only the feature of a \twonubb\ spectrum. The PSD algorithms and stability checks are being improved as analysis of the ongoing data sets are finalized. The backgrounds from the initial data sets are amongst the lowest backgrounds in a \nonubb\ region of interest and are approaching the values set by the GERDA experiment \cite{ago17}. As the analysis of the later data sets are finalized, a greater statistics background determination can be made with, at the time of writing, over 10 kg$\cdot$y of exposure. Given the expected background rate, we
project a  \nonubb\ half life sensitivity (90\%) of 10$^{26}$ y from 3-5 years of data.
At low energies, the low energy threshold and excellent energy resolution allow sensitive tests for new physics beyond the standard model. The \MJ\ success is due to its detector technology, the selection and careful handing of ultra-low-activity materials, and its low mass design. 

The combined strengths of the GERDA and \MJD\ designs are the basis for the newly formed LEGEND collaboration \cite{wil17} and its goals for a ton-scale $^{76}$Ge \nonubb\ experiment. Based on the success of the two designs, LEGEND aims to reach a background on the order of 0.1 \cpRty\ and energy resolution necessary for discovery level sensitivities in the inverted neutrino mass ordering region.

% Acknowledgement
\section{ACKNOWLEDGMENTS}
This material is based upon work supported by the U.S. Department of Energy, Office of Science, Office of Nuclear Physics, the Particle Astrophysics and Nuclear Physics Programs of the National Science Foundation, and the Sanford Underground Research Facility. 

% References

%\nocite{*}
\bibliographystyle{aipnum-cp}%
\bibliography{/Users/guiseppe/work/latex/bib/mymj}%

\end{document}